\documentclass[conference, 10pt]{IEEEtran}
\IEEEoverridecommandlockouts
\usepackage[T1]{fontenc}
\usepackage[cmintegrals]{newtxmath}
\usepackage{bm}
\usepackage[numbers, sort&compress]{natbib}
\usepackage{amsmath,amsfonts}
\usepackage{tabularx, booktabs, multirow}
\usepackage{siunitx}
\newcolumntype{Y}{>{\centering\arraybackslash}X}
\newcolumntype{Z}{>{\raggedleft\arraybackslash}X}
\newcolumntype{L}{>{\raggedright\arraybackslash}X}
\usepackage{graphicx}
\usepackage[caption=false]{subfig}
\usepackage[export]{adjustbox}
\usepackage{pgfplots}
\usepgfplotslibrary{groupplots}
\pgfplotsset{compat=1.18}
\usepackage{tikz}
\usetikzlibrary{arrows.meta, positioning, shapes.geometric, shapes, calc}
\newsavebox{\hsboxempty}
\newsavebox{\hsboxhalf}
\newsavebox{\hsboxfull}
\AtBeginDocument{%
  \sbox{\hsboxempty}{\tikz[baseline=-0.75ex]{\draw[thin](0,0)circle(.38em);}}%
  \sbox{\hsboxhalf}{\tikz[baseline=-0.75ex]{%
    \draw[thin](0,0)circle(.38em);%
    \fill(0,0)--(90:.38em)arc(90:270:.38em)--cycle;}}%
  \sbox{\hsboxfull}{\tikz[baseline=-0.75ex]{\fill(0,0)circle(.38em);}}%
}
\newcommand{\emptycirc}{\usebox{\hsboxempty}}

\newcommand{\fullcirc}{\usebox{\hsboxfull}}
\usepackage{xcolor}
\definecolor{hsurot}   {cmyk}{.00, 1, .59, .26}
\definecolor{hsugrau}  {cmyk}{.38, .37, .39, .15}
\definecolor{hsugelb}  {cmyk}{0, .16, .80, 0}
\definecolor{hsublau}  {cmyk}{1, .40, 0, .82}
\definecolor{hsuturkis}{cmyk}{1, .14, .60, .49}
\definecolor{hsugruen} {cmyk}{.16, .16, .91, .28}
\definecolor{hsuorange}{cmyk}{.01, .87, .77, .13}
\usepackage[printonlyused, nohyperlinks]{acronym}
\acrodef{pv}[PV]{photovoltaic}
\acrodef{lv}[LV]{low-voltage}
\acrodef{ev}[EV]{electric vehicle}
\acrodef{lp}[LP]{linear program}
\acrodef{milp}[MILP]{mixed-integer linear program}
\acrodef{phev}[PHEV]{plug-in hybrid electric vehicle}
\acrodef{hp}[HP]{heat pump}
\usepackage{todonotes}
\usepackage{enumitem}
\usepackage[english]{babel}
\usepackage{hyperref}
\hypersetup{colorlinks=true, unicode=true,
            linkcolor=black, citecolor=black,
            filecolor=black, urlcolor=black}
\addto\extrasenglish{}

\addto\extrasenglish{}
\addto\extrasenglish{}
\addto\extrasenglish{}
\addto\extrasenglish{}
\usepackage{titlesec}
\titlespacing*{\section}{0pt}{3pt plus 1pt minus 1pt}{2pt plus 1pt minus 1pt}
\titlespacing*{\subsection}{0pt}{2pt plus 1pt minus 1pt}{1pt plus 1pt minus 1pt}

\begin{document}
\bstctlcite{IEEEexample:BSTcontrol}
\title{Price-Optimized EV and Heat-Pump Scheduling Can Exceed Uncoordinated Feeder Peaks%
\thanks{This research in the project OptiFlex was funded by dtec.bw (Digitalization and Technology Research Center of the Bundeswehr). dtec.bw is funded by the European Union (NextGenerationEU). This work was also supported by the Federal Ministry for Economic Affairs and Energy and the Projektträger Jülich GmbH within the project InterPhaSe (FKZ 03EI6139A).}}

\author{%
\IEEEauthorblockN{Tatjana Hein\textsuperscript{1,2}, Lukas Wagner\textsuperscript{1}, Alex Bernhard\textsuperscript{1}, and Felix Ocker\textsuperscript{1}}
\IEEEauthorblockA{\textsuperscript{1}Institute of Automation Technology, Helmut Schmidt University, Hamburg, Germany \quad \textsuperscript{2}HAW Hamburg, Germany\\
tatjanasophie.hein@haw-hamburg.de,\quad \{lukas.wagner, alex.bernhard, felix.ocker\}@hsu.hamburg}
}

\maketitle

\begin{abstract}
Electric vehicles and heat pumps add large, time-flexible loads to low-voltage feeders. When each household independently minimizes its own charging cost against the same electricity price, this flexible demand piles into the cheapest hours, and the resulting synchronization can stress the feeder more than uncoordinated operation does.
This work quantifies that effect on a rural low-voltage feeder with photovoltaics (97 buses, 250\,kVA transformer) in 913 simulation runs that compare seven strategies across four seasons, device penetrations of up to 90\,\%, 208 price weeks from 2022--2025, and heterogeneous heat pumps.
At 30\,\% penetration of each device, this decentralized cost minimization drives the transformer to 253\,\% of rated capacity, more than three times the peak of non-optimized Fixed-random charging with the same mobility and charged energy, and 478\,\% with electric vehicles alone at 90\,\% penetration, with undervoltage at the far end of the longest feeder arm.
It exceeds Fixed-random and Fixed-times charging in every winter, spring, and autumn price week, and each 30\,pp of heat-pump penetration raises its peak by about 110\,pp, twelve times as much as without optimization.
Centralized peak minimization removes all demand-driven violations. In the heat-pump-free summer case at 60\,\% photovoltaics, midday reverse power flow sets a transformer floor of 128.9\,\% that no strategy undercuts, since no commuting electric vehicle is at home at the export peak.
With a time-variable grid fee added to the spot price, the weekly peak of decentralized cost minimization falls into the reduced-fee window in 99 of the 104 fee-active price weeks, compared with 53 without the fee, while keeping at least 89.8\,\% of its height, so the fee relocates rather than removes the synchronization.
The simulation and optimization code is available at \url{https://github.com/lukas-wagner/SynchronizationEffectsAnalysis}.
\end{abstract}


\section{Introduction}
\label{sec:intro}

Efforts to decarbonize the heating and transport sectors are driving large-scale adoption of \acp{hp} and \acp{ev} in residential areas. Both add substantial electrical loads to \ac{lv} feeders, and the timing of part of this consumption can be shifted within the day.
When many devices on the same feeder draw power in the same hours, their individual loads add up into a coincident peak. This synchronization can overload the distribution transformer and depress bus voltages in \ac{lv} networks~\cite{zimmermannCouplingScenariobasedGrid2026a}.

Two kinds of time-varying cost signals steer such flexible loads away from peak hours: dynamic energy prices such as intraday spot prices, and time-variable grid tariffs. Germany's \S\,14a of the Energy Industry Act (EnWG)~\cite{bundesministeriumDerJustizEnWG14a2023} is one such tariff: in exchange for accepting operator control during congestion, controllable \ac{ev} and \ac{hp} connections receive a reduced grid fee, and under its \emph{Module~3} this fee becomes a continuously time-variable tariff added to the spot price. This work models only that tariff signal, not the operator-side throttling.
A uniform cost signal, however, can re-create the peak it is meant to relieve: when every device optimizes against the same prices, demand concentrates in the cheapest window, as observed for public \ac{ev} charging under day-ahead prices~\cite{pandaAggregatePeakEV2024}. That study excludes \acp{hp} and feeder-level power flow.

Whether the same amplification arises for mixed residential \ac{ev} and \ac{hp} adoption on a rural German \ac{lv} feeder, and how price-based coordination compares there to transformer-aware scheduling, has not been systematically studied.
This work addresses that gap with a scenario-based power-flow simulation on the SimBench \texttt{1-LV-rural2-{}-0-sw} reference network, a 97-bus, 250\,kVA feeder~\cite{meineckeSimBenchBenchmarkDataset2020} whose long arms make it a demanding case for feeder voltage.
Seven scenarios, from uncontrolled heating and fixed-time charging to centralized transformer-aware optimization, are evaluated in targeted one-factor-at-a-time penetration sweeps: \ac{ev} penetration is swept over 0--90\,\%, \ac{hp} penetration over 0--60\,\% (0--90\,\% in an EV-free sweep), and \ac{pv} penetration over 0--60\,\% in 30-percentage-point (pp) steps while the others are held at a fixed reference level, across four seasonal contexts, in 160 main runs. 208 price-week realizations (728 runs) and, for one winter case, a heat-pump heterogeneity check (25 runs) test robustness, for 913 runs in total.
Three research questions guide the evaluation:

\noindent\textbf{RQ1:} How does peak transformer loading respond to increasing \ac{ev}, \ac{hp}, and \ac{pv} penetration, and how do all three indicators, transformer loading, line loading, and bus voltage, compare across seasons at a reference penetration?

\noindent\textbf{RQ2:} Where in the rural \ac{lv} feeder do synchronization-induced overloads and voltage violations concentrate spatially?

\noindent\textbf{RQ3:} Do price-based coordination strategies (intraday price, \S\,14a~EnWG Module~3) mitigate or amplify synchronization effects relative to uncoordinated charging, and what does transformer-aware optimization achieve by comparison?

The contributions are:
\begin{enumerate}[label=(\roman*), noitemsep, topsep=2pt, partopsep=0pt]
  \item A quantification of how combined \ac{ev}, \ac{hp}, and \ac{pv} penetration stresses a rural \ac{lv} feeder, separating a demand-side peak from price-induced synchronization from a summer supply-side reverse-power-flow floor that the modeled flexible devices cannot remove.
  \item Evidence that the feeder topology determines where the stress concentrates, the longest arm, while price-induced synchronization sets its severity and extent, since the end-of-line voltage violations there vanish under diversified charging start times.
  \item A systematic strategy and penetration comparison showing that price-based coordination yields the highest peak in every non-trivial demand-driven case of the main weeks and in every evaluated winter, spring, and autumn price week, that only transformer-aware peak-shaving removes the demand-driven overloads, and that the studied \S\,14a~EnWG Module\,3 grid fee relocates the synchronized peak into its reduced-fee window rather than removing it.
\end{enumerate}

\section{Related Work}
\label{sec:related}

Prior work is assessed against the problem from \autoref{sec:intro}, whether price-based coordination of combined residential loads relieves or amplifies feeder synchronization. Relevant work should jointly model \acp{ev} and \acp{hp} (R1), evaluate price-based coordination (R2) against the network-oriented control it is benchmarked on (R3), span at least two seasons (R4), test a regulatory mechanism dedicated to controllable loads such as \S\,14a~EnWG curtailment or grid fee (R5), and resolve thermal and voltage violations spatially within the feeder (R6).

One line of work models \acp{ev} and \acp{hp} together and quantifies their combined grid impact (R1).
\citet{oliyideImpactsElectricVehicles2021} project that in a typical urban British \ac{lv} network, concurrent \ac{ev} and \ac{hp} uptake drives transformer overload in an optimistic 2050 winter scenario, while cable loading and bus voltages stay within limits (R1, R4), but evaluate no coordination strategy (R2, R3), on an urban rather than the rural feeder targeted here.
\citet{tomaselliLeveragingProsumerFlexibility2024} use a rolling-horizon optimal power flow to schedule battery and thermal storage in a synthetic German \ac{lv} grid, combined with curative \ac{ev} and \ac{hp} limiting (R1, R3), and evaluate \S\,14a~EnWG curtailment (R5): storage pre-charging resolves most load violations while curtailment adds little and can worsen overloading through \ac{hp} rebound. Congestion is assessed at the HV/MV transformer rather than at individual \ac{lv} bus or line level.
\citet{zimmermannCouplingScenariobasedGrid2026a} simulate the German Federal Government energy transition pathway, with rising \ac{ev}, \ac{hp}, and \ac{pv} shares, on two SimBench reference networks under passive operation, characterizing congestion by month and by equipment rating (R1, R4). Congestion arises exclusively from transformer overloading and voltage-band violations with no line exceeding its rating, and the radial topology concentrates thermal stress at the transformer and feeder-entry lines while undervoltage concentrates at the remote feeder ends (R6).

Another line steers charging through price signals (R2).
\citet{luModellingDynamicDemand2017} propose a real-time \ac{ev} price mechanism that maximizes aggregate welfare and shaves peaks, without modeling network constraints or synchronization effects.
\citet{pandaAggregatePeakEV2024} quantify the aggregate consequence at over 1200 public charging points in the Netherlands, finding that cost-minimized charging under dynamic day-ahead prices increases aggregate peak demand by up to 16\,\% relative to unoptimized charging as charging concentrates in the same low-price period. Segmented network tariffs reduce but do not eliminate this effect. 
\citet{malkovaRecedingHorizonOptimization2024} minimize \ac{ev} charging costs at a workplace charging station under day-ahead spot prices and dynamic grid tariffs via receding-horizon \ac{milp}, enforcing a cluster power limit but performing no feeder-level power flow (R2, R4).
None of these works models \acp{hp} (R1), performs network-oriented optimization (R3), or resolves overloads at the individual bus and line level (R6).

A third line applies network-oriented control or regulatory curtailment (R3, R5).
\citet{fuReinforcementLearningBased2022} apply reinforcement learning to limit transformer overloading from uncontrolled \ac{ev} charging (R3). 
\citet{kilthauDecentralizedOptimizationApproach2025} develop a scalable decentralized energy dispatch and congestion management approach, validated on the IEEE~33- and 119-bus test networks (R3). 
\citet{galysDecentralizedAgentBasedOptimization2025} address congestion management under \S\,14a~EnWG via agent-based decentralized optimization, eliminating thermal constraint violations on the IEEE-33 test system (R3, R5). 
\citet{limmerDesignFairInterpretable2026} learn interpretable rule-based policies via multi-objective genetic programming that distribute a transformer-limited charging power among residential \acp{ev}, trading off efficiency against fairness across users (R3). 
\citet{bartaBehaviourGridorientedControlled2025} characterize \ac{hp} rebound behavior during \S\,14a~EnWG blocking and reenabling events. The response differs between inverter and fixed-speed \ac{hp}s (R5). None of these works examines price-induced synchronization peaks (R2), compares seasons (R4), or resolves overloads spatially within the feeder (R6).

\autoref{tab:related} summarizes this comparison: to the authors' knowledge, no existing work addresses all six requirements simultaneously, and neither the isolated grid effect of the Module\,3 incentive nor the point at which \ac{pv} reverse power flow overtakes demand-driven stress has been quantified.

\begin{table}[t]
  \caption{Comparison of related work against six requirements.}
  \label{tab:related}
  \centering
  \setlength{\tabcolsep}{2.5pt}
  \begin{tabular}{l cccccc}
    \toprule
    Reference & R1 & R2 & R3 & R4 & R5 & R6 \\
    \midrule
    \citet{oliyideImpactsElectricVehicles2021}             & \fullcirc  & \emptycirc & \emptycirc & \fullcirc  & \emptycirc & \emptycirc \\
    \citet{tomaselliLeveragingProsumerFlexibility2024}     & \fullcirc  & \emptycirc & \fullcirc  & \emptycirc & \fullcirc  & \emptycirc \\
    \citet{zimmermannCouplingScenariobasedGrid2026a}         & \fullcirc  & \emptycirc & \emptycirc & \fullcirc  & \emptycirc & \fullcirc  \\
    \citet{luModellingDynamicDemand2017}                   & \emptycirc & \fullcirc  & \emptycirc & \emptycirc & \emptycirc & \emptycirc \\
    \citet{pandaAggregatePeakEV2024}                       & \emptycirc & \fullcirc  & \emptycirc & \emptycirc & \emptycirc & \emptycirc \\
    \citet{malkovaRecedingHorizonOptimization2024}         & \emptycirc & \fullcirc  & \emptycirc & \fullcirc  & \emptycirc & \emptycirc \\
    \citet{fuReinforcementLearningBased2022}               & \emptycirc & \emptycirc & \fullcirc  & \emptycirc & \emptycirc & \emptycirc \\
    \citet{kilthauDecentralizedOptimizationApproach2025}   & \emptycirc & \emptycirc & \fullcirc  & \emptycirc & \emptycirc & \emptycirc \\
    \citet{galysDecentralizedAgentBasedOptimization2025}   & \emptycirc & \emptycirc & \fullcirc  & \emptycirc & \fullcirc  & \emptycirc \\
    \citet{limmerDesignFairInterpretable2026}              & \emptycirc & \emptycirc & \fullcirc  & \emptycirc & \emptycirc & \emptycirc \\
    \citet{bartaBehaviourGridorientedControlled2025}       & \emptycirc & \emptycirc & \emptycirc & \emptycirc & \fullcirc  & \emptycirc \\
    \bottomrule
  \end{tabular}
  \par\smallskip
  {\footnotesize \fullcirc~= addressed,\quad \emptycirc~= not addressed.}
\end{table}

\section{Modeling and Simulation Setup}
\label{sec:approach}

\subsection{Network and Temporal Model}

The study uses the SimBench \texttt{1-LV-rural2-{}-0-sw} reference network~\cite{meineckeSimBenchBenchmarkDataset2020}: a 97-bus rural German \ac{lv} feeder at 0.4\,kV nominal voltage with a 250\,kVA distribution transformer and 99 load connection points (92 household and 7 small-commercial).
Power flow is solved at each timestep by Newton-Raphson using pandapower.
Time is discretized at $\Delta t = \SI{0.25}{h}$ over a 7-day horizon ($T = 672$ steps), anchored to the Monday of the week containing the seasonal reference day.
Winter (calendar week~3), spring (week~16), summer (week~29), and autumn (week~42) are studied. To test robustness against the price trajectory, the reference case is additionally run for all 52 calendar weeks with the prices of 2022--2025 (208 realizations, assigned to meteorological seasons). Only the prices vary between years. Temperature, \ac{pv}, and base load keep the fixed reference profiles of the respective week.
Grid indicators are evaluated against EN\,50160 and equipment ratings: bus voltage must stay within $[0.9,\;1.1]$\,p.u., and transformer and line loading below 100\,\% of rated apparent power and rated current, respectively. Each scenario reports the peak transformer loading $\hat{S}_T$, peak line loading $\hat{l}$, minimum bus voltage $\hat{v}_{\min}$, and the hours per week the transformer and voltage limits are exceeded ($h_{T>100}$, $h_{\mathrm{UV}}$) over the 7-day horizon. Bus voltage never exceeds the upper band in any scenario (at most 1.096\,p.u.), so only undervoltage is reported. The spatial figures use the single worst-loading timestep of each scenario, defined by its peak line loading.

\subsection{Coordination Strategies}

Seven scenarios are evaluated, a baseline and six coordination strategies, as described in \autoref{tab:scenarios}.
Baseline contains only the household base load, taken from the SimBench standard load profiles, and \ac{pv} without \acp{ev} or \acp{hp}.
Both Fixed strategies charge without optimization but use the same \acp{ev} and the same trips as the optimized strategies. Each day, a charging session starts at 18:00 (Fixed-times) or at a fixed, randomly drawn time per vehicle (Fixed-random), or later when the vehicle arrives. The vehicle then charges at full wallbox power until its battery holds 45\,kWh again, following~\eqref{eq:ev_soc}. Decentralized and Decentralized-M3, which respond to the price without any grid signal, are referred to as price-based coordination, and the coincident peak they cause as price-induced synchronization.
The four optimized strategies solve \acp{lp} using Gurobi (academic licence, MIP gap 0, time limit \SI{120}{s}) over the full 7-day horizon, one independent program per device in the decentralized case and a single joint program in the centralized case.

\begin{table}[t]
  \caption{Coordination strategies evaluated in this work.}
  \label{tab:scenarios}
  \centering
  \begin{tabularx}{\columnwidth}{lL}
    \toprule
    Strategy & Description \\
    \midrule
    Baseline       & Household loads and \ac{pv} only, no \acp{ev} or \acp{hp} \\
    Fixed-times    & All \acp{ev} start at 18:00 or on later arrival at full wallbox power until 45\,kWh, \acp{hp} follow instantaneous thermal demand \\
    Fixed-random   & As Fixed-times, but each \ac{ev} starts at one time drawn from $\mathcal{N}(18{:}00,\,1.5\,\text{h})$, clipped to 14:00--22:00 and reused across the week \\
    Decentralized      & Independent \ac{lp} per device, minimizes own cost against intraday price, no grid awareness \\
    Decentralized-M3   & As Decentralized, with \S\,14a~EnWG Module\,3 grid fee added to intraday price (only applicable in Q1 and Q4, so winter and autumn simulations) \\
    Centralized-cost   & Joint \ac{lp}, minimizes total cost, transformer constraint enforced \\
    Centralized-peak   & Joint \ac{lp}, minimizes peak active-power import at the transformer \\
    \bottomrule
  \end{tabularx}
\end{table}

In the winter reference week, every strategy charges 60.6\,kWh per EV-week from the grid to replace 57.6\,kWh of trip energy at a charging efficiency of 0.95. In the other main weeks, the cost-minimizing strategies (Decentralized, Decentralized-M3, and Centralized-cost) additionally charge up to 16\,kWh at negative prices, which remains in the battery at the end of the week (up to its 60\,kWh capacity). The Fixed strategies only replace the trip energy, and every strategy serves all trips.

\subsection{Device Models}
\label{sec:devices}

\textbf{Electric vehicle.}
Each \ac{ev} $e \in \mathcal{E}$ has a 60\,kWh battery, an 11\,kW wallbox, and charging efficiency $\eta = 0.95$, with usable energy bounded by $E_{\min} = \SI{6}{kWh}$ and $E_{\max} = \SI{60}{kWh}$ (10--100\,\% state of charge) and at least the 45\,kWh start-of-week state restored by the end of the week.
Battery energy evolves as
\begin{equation}
  E_{e}(t) = E_{e}(t{-}1) + p_{e}(t)\,\eta\,\Delta t - E^{\mathrm{drv}}_{e}(t), \label{eq:ev_soc}
\end{equation}
subject to $E_{\min} \leq E_{e}(t) \leq E_{\max}$ and $0 \leq p_{e}(t) \leq P_{\text{wb}}\,a_{e}(t)$,
where $a_{e}(t)\!\in\!\{0,1\}$ is the at-home indicator and $E^{\mathrm{drv}}_{e}(t)$ is the trip energy drawn at departure.
The at-home indicator $a_{e}(t)$ is drawn from a stochastic mobility model: on weekdays a primary commute departs at $7{:}30\,\pm\,0.8\,\text{h}$ and returns at $17{:}00\,\pm\,1.5\,\text{h}$, consuming $9.0\,\pm\,2.5$\,kWh. A second evening trip ($19{:}00\,\pm\,1\,\text{h}$, $3.0\,\pm\,1.5$\,kWh) occurs with probability~0.25.
On weekends a single trip occurs with probability~0.60.
Each \ac{ev} draws independently (separate seeded stream), so plug-in times spread across the fleet. The optimized schedules keep at least 45\,kWh before every departure, whereas the Fixed strategies do not enforce this reserve, although all their trips remain feasible.

\textbf{Heat pump.}
Each \ac{hp} $h \in \mathcal{H}$ serves a single household.
Per-step thermal demand is
\begin{equation}
  Q_{h}(t) = \max\!\bigl(T_{\mathrm{thr}} - T_{\mathrm{out}}(t),\;0\bigr)\cdot\alpha\cdot\Delta t, \label{eq:hp_demand}
\end{equation}
where $T_{\mathrm{thr}} = \SI{15}{\degreeCelsius}$ is the heating threshold temperature below which space heating is required, $\alpha = \SI{0.15}{kW\per K}$ is an aggregate building heat-loss coefficient, and $T_{\mathrm{out}}(t)$ is the quarter-hourly outdoor temperature profile for Norderstedt, Germany (53.67\textdegree N, 10.02\textdegree E).
The thermal buffer $B_{h}$ evolves as
\begin{equation}
  B_{h}(t) = B_{h}(t{-}1) + \mathrm{COP}\cdot p_{h}(t)\cdot\Delta t - Q_{h}(t), \label{eq:hp_buffer}
\end{equation}
with $\mathrm{COP} = 3.5$, $P_{h,\max} = \SI{8}{kW}$, and buffer capacity $\SI{10}{kWh}$ (minimum $\SI{2}{kWh}$, with $\SI{5}{kWh}$ at the start and at least $\SI{5}{kWh}$ at the end of the week).
The 8\,kW rating markedly exceeds the per-household thermal demand, so the buffer recharges within about 20\,min. This is a deliberately homogeneous, synchronization-promoting stress-case sizing rather than a typical field design. A sensitivity case randomizes each \ac{hp} around these values with rating $\times\mathcal{N}(1,\,0.18^2)$ and COP $\times\mathcal{N}(1,\,0.19^2)$, both truncated at $\pm2\sigma$, buffer $\times\mathcal{U}[0.78,\,1.22]$, and heating threshold $+\mathcal{U}[-0.5,\,0.5]$\,K, normalized to the base mean within each of five draws. The rating, buffer, and threshold spreads follow the pool model of~\cite{fischerModelbasedFlexibilityAssessment2017}, the COP spread the field data of~\cite{brudermuellerEstimationEnergyEfficiency2025}.
At the coldest winter hour ($T_{\mathrm{out}} \approx -3\,^\circ\mathrm{C}$), the 8\,kWh usable buffer bridges about 2 to 3\,h, so the \ac{hp} can pre-heat during cheap hours and defer operation.
With this threshold, \ac{hp} thermal demand is near zero in summer (about \SI{3.6}{kWh} thermal per household-week), since only space heating and no domestic hot water is modeled, and largest in winter (about \SI{380}{kWh} thermal per household-week).

\textbf{Photovoltaic.}
\ac{pv} system capacities are drawn from $\mathcal{N}(10.0,\,3.0)$\,kWp, clipped to $[3,\,20]$\,kWp.
Generation profiles are derived from renewables.ninja capacity factors at Norderstedt.
Feed-in tariff is set to zero to ensure the priority of \ac{pv}-self-consumption in the optimization, since the grid price strictly exceeds the zero feed-in rate.

\textbf{Penetration assignment.}
\acp{ev}, \acp{hp}, and \ac{pv} systems are assigned by drawing a random permutation of the 99 load points with fixed seeds (42, 44, and 43, respectively), ensuring that the set assigned at penetration $p_1$ is a strict subset of that at $p_2 > p_1$.
This property allows effects to be attributed to individual penetration increments without confounding by load-point selection. All stochastic draws (mobility, Fixed-random start times, device placement, \ac{hp} parameters in the sensitivity case) use fixed seeds, so every scenario is exactly reproducible.

\subsection{Optimization Formulations}

Let $c(t)$ denote the electricity price, a 15-minute volume-weighted profile from the EPEX SPOT continuous intraday market for Germany (net wholesale, EUR/kWh) of 2024 for the four main weeks and of 2022--2025 for the robustness runs, mapped by ISO week and weekday, that matches the quarter-hourly control resolution used here, and $c_{\mathrm{fee}}(t)$ the Module\,3 grid fee at step $t$, taken from the Stadtwerke Norderstedt 2026 tariff schedule.
The Norderstedt grid operator applies the time-variable fee only in Q1 (January to March) and Q4 (October to December), designating 18:00--22:00 as the high-fee and 00:00--06:00 as the reduced-fee window. Outside these quarters Module\,3 provides no time-variable signal, so Decentralized-M3 is evaluated only in fee-active weeks (week~3 in Q1 and week~42 in Q4 for the main cases). In the HP-free summer PV sweep, the summer floor moreover arises when no flexible device is available (\autoref{sec:sweep}).
Let $P_{\mathrm{base}}(t)$ be the aggregate household load, $P_{\mathrm{pv}}(t)$ the total \ac{pv} output, and $P_{\mathrm{trafo}}$ the transformer power limit.

\textbf{Decentralized.}
Each device (EV or HP) solves its own \ac{lp} independently~\cite{ireshikaIEC61851Compliant2022}, minimizing its own energy cost:
\begin{equation}
  \min_{p_{e},\,p^{\mathrm{pv}}_{e}} \;\sum_{t=1}^{T} \bigl[c(t) + c_{\mathrm{fee}}(t)\bigr]\bigl[p_{e}(t) - p^{\mathrm{pv}}_{e}(t)\bigr]\Delta t, \label{eq:obj_decentral}
\end{equation}
subject to \eqref{eq:ev_soc} or \eqref{eq:hp_buffer} and device-level bounds, with $c_{\mathrm{fee}}(t)=0$ in the Decentralized scenario and the Module\,3 fee schedule in Decentralized-M3.
For \acp{ev} the availability mask sets $p_{e}(t)=0$ while the vehicle is away, and a pre-departure constraint keeps at least 45\,kWh before each departure, or more if the upcoming trip requires it.
The share $0 \leq p^{\mathrm{pv}}_{e}(t) \leq \min\bigl(p_{e}(t),\,P_{\mathrm{pv},e}(t)\bigr)$ is drawn from the household's own rooftop \ac{pv} output $P_{\mathrm{pv},e}(t)$ ($p^{\mathrm{pv}}_{e} \equiv 0$ without \ac{pv}) and is free of charge, since the feed-in tariff is zero. Feeder-wide \ac{pv} surplus is not visible to the device.
For \acp{hp} there is no availability restriction: the \ac{hp} shifts its consumption freely within the buffer bounds at any time of day or night.
No information is exchanged between households, and no transformer constraint is included. Each device acts in self-interest without grid awareness.
Since all devices share the price signal $c(t)$, synchronization emerges without explicit coordination.

\textbf{Centralized-cost.}
A single \ac{lp} coordinates all devices jointly:
\begin{equation}
  \min_{p,\,p^{\mathrm{pv}}} \;\sum_{t=1}^{T} c(t)\!\left[\sum_{e \in \mathcal{E}}\!\bigl(p_{e}(t)-p^{\mathrm{pv}}_{e}(t)\bigr)+\sum_{h \in \mathcal{H}}\!\bigl(p_{h}(t)-p^{\mathrm{pv}}_{h}(t)\bigr)\right]\!\Delta t, \label{eq:obj_central_cost}
\end{equation}
with the same zero-cost self-consumption as in~\eqref{eq:obj_decentral}. Co-located \ac{ev} and \ac{hp} self-consumption jointly cannot exceed the household's rooftop \ac{pv} output. The program is subject to device dynamics \eqref{eq:ev_soc}--\eqref{eq:hp_buffer}, device-level bounds, and
\begin{equation}
  P_{\mathrm{base}}(t) - P_{\mathrm{pv}}(t) + \sum_{e\in\mathcal{E}}p_{e}(t) + \sum_{h\in\mathcal{H}}p_{h}(t) \leq P_{\mathrm{trafo}} \quad \forall\,t. \label{eq:trafo}
\end{equation}
Constraint~\eqref{eq:trafo} bounds the active-power balance at the transformer, whereas the reported loading comes from the AC power flow and includes reactive power and losses. All \acp{ev}, \acp{hp}, and \ac{pv} systems are modeled at unity power factor, so among the modeled loads only the SimBench base load carries reactive power. A schedule that meets \eqref{eq:trafo} at the rated limit can therefore still reach about 102\,\% apparent-power loading.

\textbf{Centralized-peak.}
The objective is replaced by $\min\;\tau$, where the auxiliary variable $\tau$ upper-bounds the net feeder import at every step, and constraint~\eqref{eq:trafo} by
\begin{equation}
  P_{\mathrm{base}}(t) - P_{\mathrm{pv}}(t) + \sum_{e\in\mathcal{E}}p_{e}(t) + \sum_{h\in\mathcal{H}}p_{h}(t) \leq \tau \quad \forall\,t, \label{eq:central_peak}
\end{equation}
yielding a linear minimization of the peak active-power import at the transformer. This bounds import only, not \ac{pv} reverse power flow. \autoref{sec:sweep} shows that in the HP-free summer PV sweep no flexible device is available to absorb the export peak.

All optimizations assume perfect foresight over the 7-day horizon, an idealized common-information setting whose relaxation is discussed in \autoref{sec:conclusion}.

\section{Case Study}
\label{sec:casestudy}

\subsection{Strategy Comparison at a Reference Penetration (RQ1, RQ3)}
\label{sec:strat}

\autoref{tab:kpis} compares the grid indicators defined in \autoref{sec:approach} for each scenario at the EV=PV=HP=30\,\% reference penetration in winter, the season with the highest heat-pump-driven load.

\begin{table}[!t]
  \caption{Grid indicators for all scenarios (EV=PV=HP=30\,\%, winter). Symbols as defined in \autoref{sec:approach}.}
  \label{tab:kpis}
  \centering
  \setlength{\tabcolsep}{3pt}
  \begin{tabular}{lrrrrr}
    \toprule
    Strategy & $\hat{S}_T$ [\%] & $\hat{l}$ [\%] & $\hat{v}_{\min}$ [p.u.] & $h_{T>100}$ & $h_{\text{UV}}$ \\
    \midrule
    Baseline      & 37.9  &  28.2 & 1.008 &  0.0  & 0.0  \\
    Fixed-times   & 137.6 & 101.9 & 0.947 &  4.25 & 0.0  \\
    Fixed-random  &  78.5 &  56.3 & 0.984 &  0.0  & 0.0  \\
    Decentralized     & 253.2 & 180.1 & 0.870 &  7.0  & 1.25 \\
    Decentralized-M3  & 253.2 & 180.1 & 0.870 &  7.0  & 1.25 \\
    Centralized-cost  & 102.4 &  87.5 & 0.952 & 12.75 & 0.0  \\
    Centralized-peak  &  37.9 &  38.8 & 0.994 &  0.0  & 0.0  \\
    \bottomrule
  \end{tabular}
\end{table}

The Baseline stays well within all limits.
Starting all \acp{ev} at 18:00 (Fixed-times) raises transformer loading to 137.6\,\%, as the coincident charging start stacks onto the evening household peak (4.25\,h of overload).
Randomizing start times (Fixed-random) keeps the feeder within all limits (78.5\,\%).

Decentralized cost optimization performs substantially worse than either fixed scenario: each device independently shifts demand to the cheapest overnight hours (00:00--05:00), which concentrates load into a synchronized night-time peak at 253.2\,\% of rated capacity.
Individually rational cost minimization thus amplifies synchronization. Although the cheap window spans several hours, the linear cost minimization does not spread load across it: all devices face the same price ranking and draw full power in the same cheapest quarter-hours.
This holds across seasons: the Decentralized peak reaches 255.1\,\% in spring and 253.6\,\% in autumn, 3.8 and 3.5 times the Fixed-random peak, and 158.2\,\% in summer, where Fixed-random stays at 61\,\%. Outside winter, Decentralized also charges up to 16\,kWh per EV-week more at negative prices, which remains in the battery at the end of the week. Line loading and voltage follow the transformer: Decentralized reaches about 180\,\% line loading and 0.87\,p.u.\ in winter, spring, and autumn and still 113\,\% and 0.94\,p.u.\ in summer, whereas Fixed-random stays at or below 56\,\% and above 0.98\,p.u.\ in every season. Across all 208 price-week realizations, the winter peak ranges from 249\,\% to 263\,\% (3.1--3.7 times Fixed-random), and Decentralized exceeds both Fixed strategies in every winter, spring, and autumn realization, as does Decentralized-M3 in every fee-active one. Spring realizations span 138--256\,\% with a median of 225\,\%, so the 2024 main case in week~16 (255.1\,\%) lies near the upper end. In 16 of 52 summer realizations, Fixed-times exceeds Decentralized, and 14 of these have predominantly midday daily price minima. Fixed-random never exceeds the rating in any realization. With heterogeneous \acp{hp} (\autoref{sec:devices}), the winter Decentralized peak drops by only 1--3\,\% (246--250\,\%) and Fixed-random stays at 79\,\%, so the overload is not merely an artifact of identical \ac{hp} parameters.

Adding the Module\,3 fee (Decentralized-M3) leaves the winter peak unchanged and lowers the autumn peak by 18.4\,pp, or 7.3\,\% (\autoref{sec:module3}).

Centralized-cost coordination caps the instantaneous peak at 102.4\,\% via the transformer constraint (the excess over 100\,\% stems from reactive power and losses outside the active-power constraint~\eqref{eq:trafo}), but the optimizer fills every available hour up to that limit, yielding 12.75\,h of overload per week, whereas Fixed-random causes none.
It trades overload height for duration, saturating near 102\,\% in all four seasons with fewer overload hours when HP demand is lower.
At this reference point, Centralized-peak is the only strategy that removes all overloads, holding the transformer peak at its pre-electrification level of 37.9\,\%.

\subsection{Spatial Distribution of Overloads (RQ2)}
\label{sec:spatial}

\begin{figure}[!t]
  \centering
  \subfloat[Decentralized: worst step Wed 00:45, line max 180\,\%]{\includegraphics[width=\columnwidth]{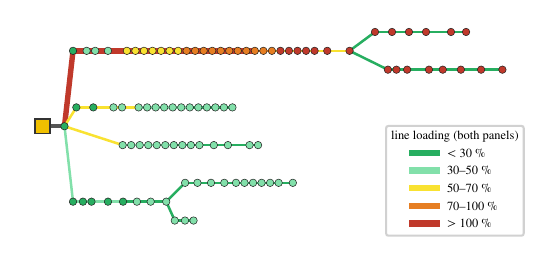}}\\[-2pt]
  \subfloat[Fixed-random: worst step Fri 18:30, line max 56\,\%]{\includegraphics[width=\columnwidth]{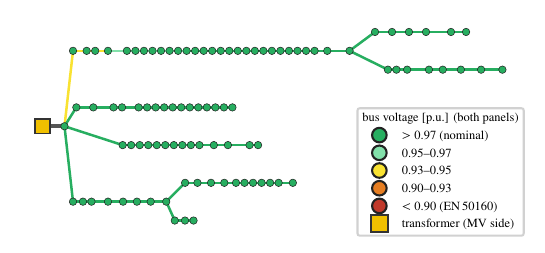}}
  \caption{Feeder loading at each strategy's peak-line-loading timestep (EV=PV=HP=30\,\%, winter). Edge color shows line loading and node color bus voltage. The horizontal position of each bus approximates its cumulative line distance from the transformer, with a minimum display spacing between adjacent buses.}
  \label{fig:network}
\end{figure}

\autoref{fig:network} contrasts the worst strategy, Decentralized coordination, with the non-optimized Fixed-random strategy, at the same reference penetration (EV=PV=HP=30\,\%, winter).

Under Decentralized coordination, overloads are not distributed uniformly: the three shorter branches connected to the transformer remain within acceptable limits, while the longest feeder arm, spanning the greatest electrical distance from the transformer, concentrates the synchronization effect.
Within that arm, line loading is highest at the first segment after the feeder split, where the full downstream load concentrates, and eases with depth as load branches off at each bus. Bus voltage, by contrast, falls cumulatively along the arm, so the end-of-line buses drop below 0.9\,p.u.\ (EN\,50160) under the synchronized load.

Under Fixed-random, the same arm carries at most 56\,\% line loading with no voltage-band violation. The synchronized overnight peak, not the feeder topology, therefore determines whether that arm actually violates its limits.

\subsection{Penetration Scaling (RQ1)}
\label{sec:sweep}

\begin{figure}[!t]
  \centering
  \includegraphics[width=0.92\columnwidth]{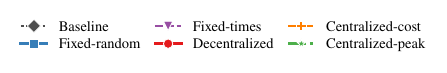}\\[-1pt]
  \subfloat[HP penetration (\ac{ev}=30\,\%, PV=0\,\%, winter)\label{fig:sweep_hp}]{\includegraphics[width=\columnwidth]{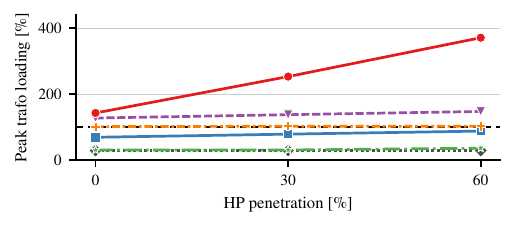}}\\[-2pt]
  \subfloat[PV penetration (\ac{ev}=30\,\%, HP=0\,\%, summer)\label{fig:sweep_pv}]{\includegraphics[width=\columnwidth]{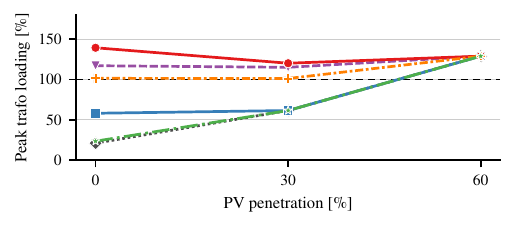}}
  \caption{Peak transformer loading versus device penetration. Decentralized-M3 is omitted: its peak coincides with Decentralized in winter, and the time-variable fee does not apply in summer.}
  \label{fig:sweep}
\end{figure}

\autoref{fig:sweep} isolates the incremental effect of \acp{hp} and \ac{pv}: rising \ac{hp} penetration in winter (\autoref{fig:sweep_hp}) and rising PV penetration in summer (\autoref{fig:sweep_pv}), both at the 30\,\% \ac{ev} reference level from \autoref{sec:strat}. The \ac{ev}-only sweep (HP=PV=0\,\%, winter) is discussed first, from the same runs.

Under Decentralized coordination and without \acp{hp} and \ac{pv}, \ac{ev} penetration drives a steep, monotonic increase in peak loading, going from 143\,\% at 30\,\% \acp{ev} penetration to 478\,\% at 90\,\% \acp{ev} penetration.
Fixed-random scales more gradually (152\,\% at 90\,\% \ac{ev}).
Fixed-times rises to 369\,\% at 90\,\% \ac{ev} but stays below Decentralized at every positive \ac{ev} penetration (127 vs.\ 143\,\% at 30\,\%).
Centralized-cost coordination caps the peak near 102\,\% across all \ac{ev} penetration levels, but overload duration grows from 6.5\,h (\ac{ev}=30\,\%) to 22.0\,h (\ac{ev}=90\,\%). Additional load converts to longer overload rather than higher peaks, because the hard cap fixes the peak height and extra demand can only fill more hours. Centralized-peak holds transformer loading near the baseline level, between 30 and 32\,\% across all \ac{ev} levels.

\ac{hp} scaling (\autoref{fig:sweep_hp}) widens this gap, since price-responsive \acp{hp} shift additional load to the same overnight cheap window as the \acp{ev}.
Under Decentralized coordination, each 30\,pp increment of HP penetration adds approximately 110\,pp to the transformer peak: from 143\,\% (HP=0\,\%) to 253\,\% (HP=30\,\%) to 371\,\% (HP=60\,\%).
Under fixed scheduling, the same increments add only about 10\,pp per step: Fixed-random rises from 69\,\% to 88\,\%, and Fixed-times from 127\,\% to 147\,\%.
Since all strategies serve the same space-heating demand and differ only marginally in terminal buffer energy, the much larger increment is attributable to \ac{hp} price-responsiveness rather than heat-demand volume.
Centralized-cost coordination again saturates at 102\,\% peak regardless of HP penetration, but overload duration approximately doubles with each HP increment: at PV=0\,\% it grows from 6.5\,h (HP=0\,\%) to 14.5\,h (HP=30\,\%) and 28.0\,h (HP=60\,\%). The longest overload of all runs, 30.5\,h with minimum bus voltage 0.94\,p.u., occurs at EV=90\,\% with PV=HP=30\,\%. The 14.5\,h exceed the 12.75\,h of \autoref{tab:kpis} because the table fixes PV=30\,\%.
Centralized-peak holds transformer loading below 36\,\% across the swept HP levels.

In winter (\ac{ev}=30\,\%, HP=0\,\%), \ac{pv} leaves the peaks of Decentralized (143\,\%) and Fixed-times unchanged, and Centralized-cost stays near 101\,\%. For Baseline, Centralized-peak, and Fixed-random, midday export becomes peak-setting at PV=60\,\% (88\,\%).
In summer, where \ac{pv} production is substantial, rising \ac{pv} penetration replaces overnight demand as the dominant source of transformer stress (\autoref{fig:sweep_pv}).
At \ac{pv}=0\,\%, the strategy spread mirrors the winter peak-pattern at reduced scale.
As \ac{pv} penetration grows, midday reverse power flow progressively compresses this spread.
At \ac{pv}=60\,\%, all strategies converge to a common floor of 128.9\,\%, the peak reverse power flow from \ac{pv} export. At this export peak (a weekday at 11:00) none of the 30 \acp{ev} is at home, since the mobility model sends every \ac{ev} on a weekday commute, and no \ac{hp} is installed, so no modeled flexible device can absorb the export. Lowering the floor requires curtailment or storage.
The convergence illustrates a qualitative limit: once transformer overloading is supply-driven rather than demand-driven, demand-side coordination loses leverage.

\subsection{{§}\,14a~EnWG Module\,3 Grid Fee (RQ3)}
\label{sec:module3}

The Module\,3 fee is active only in Q1 and Q4, covering the studied winter and autumn weeks. Adding it to Decentralized coordination leaves the peak unchanged in winter at every tested penetration (143\,\%, 287\,\%, and 478\,\% at 30, 60, and 90\,\% \ac{ev} with HP=0\,\%), because the fee's 00:00--06:00 reduced-fee window coincides with the cheapest intraday hours. In autumn a 4--5\,pp reduction emerges across the \ac{ev} sweep, as the night-time price minimum is less pronounced in week~42 and the fee gains a marginal incentive to shift demand. The larger 18.4\,pp drop at the mixed reference penetration (\autoref{sec:strat}) reflects its higher baseline peak, so in relative terms every fee effect stays below 8\,\%.

These results are specific to the studied weeks. Over the Q1/Q4 days of 2022--2025, the daily price minimum falls into the reduced-fee window on only 40\,\% of days, and the fee changes the timing of the cheapest quarter-hour on 32\,\% of days. Across the 104 fee-active week realizations of the reference case, the fee leaves the peak exactly unchanged in 39 weeks and within 1\,pp in 51. It lowers the peak by more than 1\,pp in 44 weeks (by up to 25.1\,pp, 10.2\,\%) and raises it in 9 (by up to 25.6\,pp, 11.4\,\%). With the fee, the weekly transformer maximum falls into the reduced-fee window between 00:00 and 06:00 in 99 of the 104 weeks, compared with 53 weeks without the fee, and it stays at no less than 89.8\,\% of its value without the fee. The studied fee thus relocates the synchronized demand rather than removing it.

The HP scaling (EV=0\,\%) shows no reduction in either season: \acp{hp} already concentrate all shiftable thermal demand into the cheapest overnight hours through the spot price signal, so their buffer constraints leave no residual flexibility for the Module\,3 fee to exploit.
Without a channel to feeder state, as in the agent-based scheme of \citet{galysDecentralizedAgentBasedOptimization2025}, the studied fee alone does not deliver comparable relief.

\section{Conclusion}
\label{sec:conclusion}

This work quantified how a rural \ac{lv} feeder responds when \acp{ev} and \acp{hp} follow a shared price signal, and how far coordinated scheduling and a regulated tariff relieve the resulting stress.

\textbf{RQ1:} At the reference penetration (EV=HP=PV=30\,\%), decentralized cost minimization drives the transformer to 253\,\% of rated capacity, more than three times the energy-equivalent Fixed-random benchmark, which stays within all limits. With \acp{ev} alone at 90\,\% penetration, it reaches 478\,\%. \acp{hp} amplify this disproportionately, adding about 110\,pp per 30\,pp of HP penetration, about twelve times the roughly 10\,pp per increment under fixed scheduling, for this feeder and the homogeneous HP sizing. The mechanism holds for all three indicators across winter, spring, and autumn. For transformer loading, the 52 winter price realizations span 249--263\,\%. In the HP-free summer PV sweep, PV reverse power flow instead sets a 128.9\,\% transformer floor that the modeled flexible devices cannot lower.

\textbf{RQ2:} The overloads concentrate on the longest feeder arm. Line loading peaks at its entry, where the full downstream load concentrates, while cumulative impedance drives the end-of-line voltages below 0.9\,p.u. Shorter branches stay within limits.

\textbf{RQ3:} Price-based coordination yields the highest peak in every non-trivial demand-driven case of the main weeks and exceeds both Fixed strategies in all evaluated winter, spring, and autumn price realizations. Fixed-times exceeds it in 16 of 52 summer realizations.
Centralized-cost coordination holds the apparent-power peak near 102\,\%, since its cap acts on active power, but stretches the overload over time (up to 30.5\,h per week) and leaves less voltage margin (down to 0.94\,p.u.), whereas Centralized-peak removes all demand-driven violations, limited only by the supply-driven PV floor.
In the main weeks, Module\,3 leaves the winter reference peak unchanged and lowers the autumn peaks by 4--5\,pp in the EV sweep and by 18.4\,pp (7\,\%) at the reference. Across 104 fee-active price weeks, the studied fee relocates price-induced synchronization rather than removing it: the weekly transformer peak falls into the reduced-fee window in 99 weeks, at no less than 89.8\,\% of its height without the fee, and the fee even raises the peak by more than 1\,pp in 9 weeks.
In this study, only coordination with an explicit grid constraint reliably removed the violations, which supports grid-coupled schemes such as~\cite{kilthauDecentralizedOptimizationApproach2025}. Heterogeneous \ac{hp} parameters alone barely desynchronize devices that share one price signal, and segmented network tariffs only reduce the effect~\cite{pandaAggregatePeakEV2024}.

Several limitations bound these findings. All optimizations assume perfect foresight over the 7-day horizon, whereas practical implementations must operate under forecast uncertainty.
Five normalized \ac{hp}-parameter draws for one winter case lower the peak by only 1--3\,\%, but \ac{ev} hardware and building coefficients remain homogeneous, the COP is constant, and heat demand covers space heating only, without domestic hot water, so the \ac{hp}-driven results apply to the heating season. Outside winter, the cost-minimizing strategies charge up to 16\,kWh per EV-week more at negative prices, which remains in the battery at the end of the week but correlates only weakly with the weekly peaks across the 208 realizations. The mobility model sends every \ac{ev} on a weekday commute, without home office or second cars. Fixed-random rests on one draw of start times, and both Fixed strategies skip the 45\,kWh pre-departure reserve.
The study uses a single reference network topology, and the Module\,3 results hold for the Norderstedt fee schedule. Only prices vary between price years, while temperature, PV, and base load are fixed reference profiles.

Future work should examine receding-horizon control under forecast uncertainty, time-integrated overload-duration constraints, a fair allocation of the capped feeder capacity among households~\cite{limmerDesignFairInterpretable2026}, whether other time-variable tariffs relocate synchronization in the same way, and additional feeder topologies.

\bibliographystyle{IEEEtranN}
\scriptsize{\bibliography{literature}}

\begin{thebibliography}{16}
\providecommand{\natexlab}[1]{#1}
\providecommand{\url}[1]{#1}
\csname url@samestyle\endcsname
\providecommand{\newblock}{\relax}
\providecommand{\bibinfo}[2]{#2}
\providecommand{\BIBentrySTDinterwordspacing}{\spaceskip=0pt\relax}
\providecommand{\BIBentryALTinterwordstretchfactor}{4}
\providecommand{\BIBentryALTinterwordspacing}{\spaceskip=\fontdimen2\font plus
\BIBentryALTinterwordstretchfactor\fontdimen3\font minus
  \fontdimen4\font\relax}
\providecommand{\BIBforeignlanguage}[2]{{%
\expandafter\ifx\csname l@#1\endcsname\relax
\typeout{** WARNING: IEEEtranN.bst: No hyphenation pattern has been}%
\typeout{** loaded for the language `#1'. Using the pattern for}%
\typeout{** the default language instead.}%
\else
\language=\csname l@#1\endcsname
\fi
#2}}
\providecommand{\BIBdecl}{\relax}
\BIBdecl

\bibitem[Zimmermann et~al.(2026)Zimmermann, Wagner, {von R{\"o}nn}, Strobel,
  H{\"u}ttmann, and Gehlhoff]{zimmermannCouplingScenariobasedGrid2026a}
N.~Zimmermann \emph{et~al.}, ``Coupling scenario-based grid simulations with
  state estimation: {{Measurement}} requirements for low-voltage networks under
  the german energy transition pathway,'' \emph{Energies}, vol.~19, no.~15, p.
  3494, 2026.

\bibitem[{Bundesministerium der
  Justiz}(2023)]{bundesministeriumDerJustizEnWG14a2023}
{Bundesministerium der Justiz}, ``{Energiewirtschaftsgesetz (EnWG) {§}\,14a:
  Netzorientierte Steuerung steuerbarer Verbrauchseinrichtungen},'' 2023.

\bibitem[Panda et~al.(2024)Panda, Li, and Tindemans]{pandaAggregatePeakEV2024}
N.~K. Panda \emph{et~al.}, ``Aggregate {{Peak EV Charging Demand}}: {{The
  Influence}} of {{Segmented Network Tariffs}},'' in \emph{2024 {{IEEE
  Transportation Electrification Conference}} and {{Expo}}}.\hskip 1em plus
  0.5em minus 0.4em\relax IEEE, 2024, pp. 1--6.

\bibitem[Meinecke et~al.(2020)Meinecke, Sarajli\'c, Drauz, Klettke, Lauven,
  Rehtanz, Moser, and Braun]{meineckeSimBenchBenchmarkDataset2020}
S.~Meinecke \emph{et~al.}, ``{{SimBench}}---{{A Benchmark Dataset}} of
  {{Electric Power Systems}} to {{Compare Innovative Solutions Based}} on
  {{Power Flow Analysis}},'' \emph{Energies}, vol.~13, no.~12, p. 3290, 2020.

\bibitem[Oliyide and Cipcigan(2021)]{oliyideImpactsElectricVehicles2021}
R.~O. Oliyide and L.~M. Cipcigan, ``The impacts of electric vehicles and heat
  pumps load profiles on low voltage distribution networks in {{Great Britain}}
  by 2050,'' \emph{International Multidisciplinary Research Journal}, pp.
  30--45, 2021.

\bibitem[Tomaselli et~al.(2024)Tomaselli, Most, Sinani, Stursberg, Heger, and
  Niessen]{tomaselliLeveragingProsumerFlexibility2024}
D.~Tomaselli \emph{et~al.}, ``Leveraging {{Prosumer Flexibility}} to {{Mitigate
  Grid Congestion}} in {{Future Power Distribution Grids}},'' \emph{Energies},
  vol.~17, no.~17, p. 4217, 2024.

\bibitem[Lu et~al.(2017)Lu, Qi, Zhang, He, and
  Zhao]{luModellingDynamicDemand2017}
Z.~Lu \emph{et~al.}, ``Modelling dynamic demand response for plug-in hybrid
  electric vehicles based on real-time charging pricing,'' \emph{IET
  Generation, Transmission \& Distribution}, vol.~11, no.~1, pp. 228--235,
  2017.

\bibitem[Malkova et~al.(2024)Malkova, Zepter, Marinelli, Amezquita, and
  Morais]{malkovaRecedingHorizonOptimization2024}
A.~Malkova \emph{et~al.}, ``Receding horizon optimization for distributed
  control of electric vehicle charging stations,'' in \emph{Proceedings of
  {{IEEE PES ISGT Europe}} 2024}.\hskip 1em plus 0.5em minus 0.4em\relax
  Dubrovnik, Croatia: IEEE, 2024.

\bibitem[Fu et~al.(2022)Fu, Schulz, Versen, and
  Stiemer]{fuReinforcementLearningBased2022}
Y.~Fu \emph{et~al.}, ``Reinforcement learning based charging management for
  electric vehicle fleet to reduce transformer overloading in distribution
  networks,'' in \emph{dtec.bw-{{Beitr\"age}} der
  {{Helmut-Schmidt-Universit\"at}}}, 2022.

\bibitem[Kilthau et~al.(2025)Kilthau, Henkel, Wagner, Gehlhoff, and
  Fay]{kilthauDecentralizedOptimizationApproach2025}
M.~Kilthau \emph{et~al.}, ``A decentralized optimization approach for scalable
  agent-based energy dispatch and congestion management,'' \emph{Applied
  Energy}, vol. 377, p. 124606, 2025.

\bibitem[Galys et~al.(2025)Galys, Wagner, Kilthau, Fay, and
  Gehlhoff]{galysDecentralizedAgentBasedOptimization2025}
J.~Galys \emph{et~al.}, ``Decentralized {{Agent-Based Optimization}} for
  {{Congestion Management}} in {{Low-Voltage Grids Incorporating Legally
  Mandated}} and {{Voluntary Flexibility}},'' in \emph{{{IET Conference
  Proceedings}}}, no.~45.\hskip 1em plus 0.5em minus 0.4em\relax IET, 2025, pp.
  326--333.

\bibitem[Limmer et~al.(2026)Limmer, Kenny, Ray, Lanfermann, Singh, and
  Castellani]{limmerDesignFairInterpretable2026}
S.~Limmer \emph{et~al.}, ``Design of fair and interpretable electric vehicle
  charging policies through genetic programming,'' \emph{Applied Energy}, vol.
  404, p. 127176, 2026.

\bibitem[Barta et~al.(2025)Barta, Gebert, Baumgartner, Uhrig, and
  Witzmann]{bartaBehaviourGridorientedControlled2025}
V.~Barta \emph{et~al.}, ``Behaviour of grid-oriented controlled heat pumps,''
  in \emph{IET Conference Proceedings}, vol. 2024, no.~5.\hskip 1em plus 0.5em
  minus 0.4em\relax IET, 2025, pp. 459--463.

\bibitem[Fischer et~al.(2017)Fischer, Wolf, Wapler, Hollinger, and
  Madani]{fischerModelbasedFlexibilityAssessment2017}
D.~Fischer \emph{et~al.}, ``Model-based flexibility assessment of a residential
  heat pump pool,'' \emph{Energy}, vol. 118, pp. 853--864, 2017.

\bibitem[Brudermueller et~al.(2025)Brudermueller, Potthoff, Fleisch, Wortmann,
  and Staake]{brudermuellerEstimationEnergyEfficiency2025}
T.~Brudermueller \emph{et~al.}, ``Estimation of energy efficiency of heat pumps
  in residential buildings using real operation data,'' \emph{Nature
  Communications}, vol.~16, no.~1, p. 2834, 2025.

\bibitem[Ireshika and Kepplinger(2022)]{ireshikaIEC61851Compliant2022}
M.~Ireshika and P.~Kepplinger, ``{{IEC}} 61851 compliant demand side management
  algorithm for electric vehicle charging: A {{MILP}} based decentralized
  approach,'' in \emph{13th {{Mediterranean Conference}} on {{Power
  Generation}}, {{Transmission}}, {{Distribution}} and {{Energy
  Conversion}}}.\hskip 1em plus 0.5em minus 0.4em\relax Valletta, Malta: IET,
  2022, pp. 146--152.

\end{thebibliography}

\end{document}